\documentclass[10pt, conference, compsocconf]{IEEEtran}
\usepackage{amsmath}
\usepackage{amssymb}
\usepackage{cite}
\usepackage{algorithm}
\usepackage{booktabs}
\usepackage{verbatim} 
\usepackage{subfigure}
\usepackage{balance}
\usepackage{graphicx}
\usepackage{tikz}
\usepackage{algpseudocode}
\usepackage{url}
\usetikzlibrary{fit,positioning}

\usepackage{cite}
\IEEEoverridecommandlockouts

\ifCLASSINFOpdf

\else
\fi

\begin{document}
	%
	\title{Poster Abstract: A Dynamic Data-Driven Prediction Model for Sparse Collaborative Sensing Applications}
	\author{\IEEEauthorblockN{Daniel (Yue) Zhang*, Yang Zhang*\thanks{*The first two authors contributed equally to this work.}, Dong Wang}

\IEEEauthorblockA{
University of Notre Dame, IN, USA\\
yzhang40@nd.edu, yzhang42@nd.edu, dwang5@nd.edu}}
	\maketitle
	%
	\IEEEpeerreviewmaketitle
	\begin{abstract}
A fundamental problem in collaborative sensing lies in providing an accurate prediction of critical events (e.g., hazardous environmental condition, urban abnormalities, economic trends). However, due to the resource constraints, collaborative sensing applications normally only collect measurements from a subset of physical locations and predict the measurements for the rest of locations. This problem is referred to as \emph{sparse collaborative sensing prediction}.  In this poster, we present a novel closed-loop prediction model by leveraging topic modeling and online learning techniques. We evaluate our scheme using a real-world collaborative sensing dataset. The initial results show that our proposed scheme outperforms the state-of-the-art baselines.  
\end{abstract}

\section {Introduction}
Collaborative sensing has emerged as a new sensing paradigm where measurements about the physical environment are collected from a set of collaborative sensors (e.g., sensing stations, vehicles, crowds) \cite{wang2018age, zhang2017towards, wang2013credibility}. An important task in collaborative sensing is to predict physical measurements based on a limited amount of collected sensory data (e.g., forecasting health risks, traffic delays, economic trends) \cite{wang2013recursive}. This prediction task is challenging due to three reasons. First,  collaborative sensing applications normally only collect data from a subset of locations due to resource limitations (e.g., limited amount of sensors). The missing data will naturally present in both spatial and temporal dimensions \cite{zhang2018robust}. 
Second, the physical variable to be predicted can be highly dynamic by nature. It is often difficult to accurately capture the large spatiotemporal dynamics of those variables in real-time \cite{zhang2018risksens}. Third, the prediction results are directly affected by the task allocation strategy that decides when and where to collect the measurements in the sensing process \cite{zhang2018optimizing}. However, it is not a trivial task to design a closed-loop system that can automatically adjust the task allocation strategy to optimize the overall prediction accuracy. A solution that can effectively address the above challenges has yet to be developed.

\section {Solution}
In this paper, we present a novel prediction model to address the above challenges. Specifically, our scheme jointly answers two questions: 1) how to accurately predict future measurements of all locations given sensing data from a sparse subset of locations? 2) how to decide which locations to collect data in order to minimize prediction error? This scheme is inspired by Dynamic Data-Driven Applications Systems (DDDAS)~\cite{madey2007enhanced}. We discuss the technical details below.



\subsection{Prediction with  Dynamic Spatio-Temporal Variation}
We first develop a new state space autoregressive model combined with dynamic topic modeling to predict future measurements of all locations. To handle spatio-temporal variation of the data. We designed a Dynamic Spatial-constrained Autoregressive (DSAR) model to infer future data based on both spatial and temporal correlations. An  order-p DSAR model is formulated as:
\begin{equation}\label{eq:model}
\small
x(t) =  \sum_{i=1}^{p} W_i^t \phi_i x(t-i) + u_t
\end{equation}\noindent
where the autoregressive term $\sum_{i=1}^{p} W_i^t \phi_i x(t-i)$ allows the model to capture the temporal dependency over time.  $W_i^t$ is a $S \times S$ dynamic spatial weight matrix that models the ``distance" between locations. Each element in the matrix represents the strength of dependency between two locations.  $\phi$ is a $S \times S$ diagonal space-time transition matrix at lag $i$.  $u_t$ is the $S \times 1$ white noise with covariance matrix Q.  

The dynamic spatial weight matrix $W_i^t$ allows the model to infer unseen data from spatial neighbors. We assume spatial correlations to be highly dynamic and can not be simply inferred from the physical distance or some prior knowledge. To infer spatial correlation, we leverage the topic modeling technique, which is originally used to infer the hidden ``topics" in text documents. Specifically, we consider the measurements (a time series) at each location are characterized by various latent features (i.e. ``topics").  The spatial correlation thus can be regarded as the similarities among latent features. We leverage a temporal-aware topic model (TTM) to dynamically infer such latent features \cite{zhang2018robust}.

\subsection{Closed-loop Task Allocation}
We further develop a closed-loop task allocation scheme to select the optimal $k$ locations in each sensing cycle to collect measurements.  The ultimate goal of this scheme is to improve the accuracy of the prediction algorithm by recommending the locations that yield the lowest prediction errors. 


To find the optimal task assignment, we first sort the priority of the locations to be sensed based on three different factors, namely, \emph{temporal uncertainty (TU)},  \emph{inference freshness (IF)}, and \emph{alertness (AT)}.  Here, TU refers to the variance of the set of predicted values in different cycles for each location. IF indicates the waiting time for each location that not been selected for sensing. AT means the implication of hazard severity induced by the predicted value. Formally, we order the priority of locations to deploy sensors to as: 


\begin{equation}\label{eq:model}
\small
Order^t =  \lambda_1^t TU + \lambda_2^t IF + \lambda_3^t AT 
\end{equation}

While these factors determine how we should pick the optimal task assignment, the weight of each factor needs to be dynamically updated due to the variation of both incoming data and the performance of the prediction algorithm. 

We propose an online learning method called \textit{EWIEM (Exponential Weights-Based-Inference-Error-Minimization)} that aims to minimize the prediction error and rank locations by combining the three factors list above, and adjust the weights dynamically. The key idea of the algorithm is to provide feedback (defined as a loss function $\ell_i$) for each expert and update expert weight accordingly. 
 Formally, we compute: $\ell_i = RMSE(Order^t,k) \cdot sim(Order^t,Order_i^t)$, where  $Order_i^t$ is the order given by each individual expert (i.e.,TU, IF, AT). $sim(\cdot)$ is a function that calculates the similarity between two order sequences: $sim(seqA,seqB,k) = \sum_{i=1}^{k} i* (seqA[i]==seqB[i]?1:0)$. 

%

				\section{Results}

We use a real world colalborative sensing dataset in \cite{zhang2018robust} to evaluate our DSAR scheme. The dataset contains hourly air quality measurements from 12 monitoring stations in Beijing, China. We use this dataset to evaluate the effectiveness of our proposed scheme in predicting hazardous air quality. 


We first evaluate the performance of DSAR in terms of Root Mean Square Error (RMSE) compared to a few representative predictive models.  The details of the baselines are given in \cite{zhang2018robust}. We randomly remove a subset of data (10\% to 90\%) from the original dataset to simulate a sparse collaborative sensing scenario. We observe that the DSAR scheme outperforms others in both dense and sparse settings (Table \ref{tab:accuracyo2}).


\begin{table}[!thpb]
		\vspace{-0.1in}
	\scriptsize
	\caption{Prediction Error (RMSE)  w/ varying Sparsity}
	\centering
	\begin{tabular}{|l|l|l|l|l|l|l|}
		\hline
		&    Sparsity=0     &0.1 &0.3& 0.5 &0.7&0.9 \\ \hline
		DSAR      & 49.41   & 57.49
		&61.18&66.44&71.05&81.67\\ \hline
		STARMAX          & 65.92 & 72.54&84.83
		&89.71&99.02&115.34 \\\hline
		ARIMA       & 63.41
		&65.20&70.89&77.27&83.02  &
		89.12 \\\hline
		SVR     & 102.82   & 126.18
		&151.09&165.09&185.24&197.93\\ \hline
		KKF      & 58.87 &64.36&75.28&91.68&93.77&96.63  \\\hline
		
	\end{tabular}
			\vspace{-0.1in}

	\label{tab:accuracyo2}
\end{table}



\begin{figure}[!htb]
\vspace{-0.05in}
	\centering
	\includegraphics[width=6.5cm]{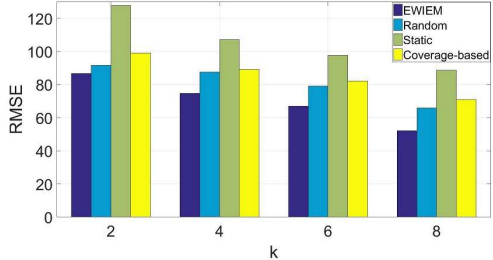}
		\vspace{-0.1in}

	\caption{Comparing Task Allocation Strategies vs. Prediction Error}
	\label{fig:ewiem}
	\vspace{-0.15in}
\end{figure}

We then evaluate the effectiveness of our optimal task allocation algorithm using the feedback control - EWIEM.  We compare EWIEM with three heuristic approaches - \emph{Random}, \emph{Static} and \emph{Coverage-based} \cite{zhang2018optimizing}, while varying the number of sensors $k$ (Figure~\ref{fig:ewiem}). We observe that our approach achieves the least the prediction error by using the proposed feedback control algorithm. 

\section{Conclusion}

This poster presents a new data-driven solution for sparse collaborative sensing applications. It incorporates a prediction algorithm that can effectively capture spatiotemporal dynamics of the measured varialbes and a closed-loop task allocation mechanism to optimize prediction performance under budget constraints. Results on a real-world case study show the proposed scheme outperforms baselines.

\section*{Acknowledgments}
This research is supported in part by the National Science Foundation under Grant No. CNS-1831669, CBET-1637251, CNS-1566465 and IIS-1447795, Army Research Office under Grant W911NF-17-1-0409. The views and conclusions contained in this document are those of the authors and should not be interpreted as representing the official policies, either expressed or implied, of the Army Research Office or the U.S. Government. The U.S. Government is authorized to reproduce and distribute reprints for Government purposes notwithstanding any copyright notation here on.

\bibliographystyle{IEEEtran}
\bibliography{refs} 

\begin{thebibliography}{1}
\providecommand{\url}[1]{#1}
\csname url@samestyle\endcsname
\providecommand{\newblock}{\relax}
\providecommand{\bibinfo}[2]{#2}
\providecommand{\BIBentrySTDinterwordspacing}{\spaceskip=0pt\relax}
\providecommand{\BIBentryALTinterwordstretchfactor}{4}
\providecommand{\BIBentryALTinterwordspacing}{\spaceskip=\fontdimen2\font plus
\BIBentryALTinterwordstretchfactor\fontdimen3\font minus
  \fontdimen4\font\relax}
\providecommand{\BIBforeignlanguage}[2]{{%
\expandafter\ifx\csname l@#1\endcsname\relax
\typeout{** WARNING: IEEEtran.bst: No hyphenation pattern has been}%
\typeout{** loaded for the language `#1'. Using the pattern for}%
\typeout{** the default language instead.}%
\else
\language=\csname l@#1\endcsname
\fi
#2}}
\providecommand{\BIBdecl}{\relax}
\BIBdecl

\bibitem{wang2018age}
D.~Wang, B.~K. Szymanski, T.~Abdelzaher, H.~Ji, and L.~Kaplan, ``The age of
  social sensing,'' \emph{IEEE Computer}, 2018.

\bibitem{zhang2017towards}
D.~Zhang, C.~Zheng, D.~Wang, D.~Thain, X.~Mu, G.~Madey, and C.~Huang, ``Towards
  scalable and dynamic social sensing using a distributed computing
  framework,'' in \emph{2017 IEEE 37th International Conference on Distributed
  Computing Systems (ICDCS)}.\hskip 1em plus 0.5em minus 0.4em\relax IEEE,
  2017, pp. 966--976.

\bibitem{wang2013credibility}
D.~Wang, L.~Kaplan, T.~Abdelzaher, and C.~C. Aggarwal, ``On credibility
  estimation tradeoffs in assured social sensing,'' \emph{IEEE Journal on
  Selected Areas in Communications}, vol.~31, no.~6, pp. 1026--1037, 2013.

\bibitem{wang2013recursive}
D.~Wang, T.~Abdelzaher, L.~Kaplan, and C.~C. Aggarwal, ``Recursive
  fact-finding: A streaming approach to truth estimation in crowdsourcing
  applications,'' in \emph{2013 IEEE 33rd International Conference on
  Distributed Computing Systems}.\hskip 1em plus 0.5em minus 0.4em\relax IEEE,
  2013, pp. 530--539.

\bibitem{zhang2018robust}
D.~Zhang, Y.~Zhang, Q.~Li, N.~Vance, and D.~Wang, ``Robust state prediction
  with incomplete and noisy measurements in collaborative sensing,'' in
  \emph{2018 IEEE 15th International Conference on Mobile Ad Hoc and Sensor
  Systems (MASS)}.\hskip 1em plus 0.5em minus 0.4em\relax IEEE, 2018, pp.
  460--468.

\bibitem{zhang2018risksens}
Y.~Zhang, Y.~Lu, D.~Zhang, L.~Shang, and D.~Wang, ``Risksens: A multi-view
  learning approach to identifying risky traffic locations in intelligent
  transportation systems using social and remote sensing,'' in \emph{2018 IEEE
  International Conference on Big Data (Big Data)}.\hskip 1em plus 0.5em minus
  0.4em\relax IEEE, 2018, pp. 1544--1553.

\bibitem{zhang2018optimizing}
Y.~Zhang, D.~Zhang, N.~Vance, and D.~Wang, ``Optimizing online task allocation
  for multi-attribute social sensing,'' in \emph{2018 27th International
  Conference on Computer Communication and Networks (ICCCN)}.\hskip 1em plus
  0.5em minus 0.4em\relax IEEE, 2018, pp. 1--9.

\bibitem{madey2007enhanced}
G.~Madey, A.-L. Barab{\'a}si, N.~Chawla, M.~Gonzalez, D.~Hachen, B.~Lantz,
  A.~Pawling, T.~Schoenharl, G.~Szab{\'o}, P.~Wang \emph{et~al.}, ``Enhanced
  situational awareness: Application of dddas concepts to emergency and
  disaster management,'' \emph{Computational Science--ICCS 2007}, pp.
  1090--1097, 2007.

\end{thebibliography}
\end{document}